\newcommand{\be}{\begin{equation}}\newcommand{\ee}{\end{equation}}
\newcommand{\bea}{\begin{eqnarray}}\newcommand{\eea}{\end{eqnarray}}
\newcommand{\ba}{\begin{array}}\newcommand{\ea}{\end{array}}
\newcommand{\mult}[3]{{\bf(#1,#2,#3)}}

\newcommand{\nn}{\nonumber}
\newcommand{\p}[1]{(\ref{#1})}
\newcommand{\bD}{\bar D}
\renewcommand{\th}{\theta}
\newcommand{\bth}{{\bar\th}}
\newcommand{\bpsi}{{\bar\psi}}
\newcommand{\bPsi}{{\overline\Psi}}
\newcommand{\bPhi}{{\overline\Phi}}
\newcommand{\Lam}{\Lambda}
\newcommand{\bLam}{\bar\Lam}
\newcommand{\eps}{{\epsilon}}
\newcommand{\one}{{\it{1}}}

\newcommand{\diff}{\mbox{d}\,}

\documentclass[12pt]{article}
\usepackage{amscd,amsmath,amssymb}

\begin{document}

\thispagestyle{empty}
\vspace{2cm}
\begin{flushright}
hep-th/0604215 \\[3cm]
\end{flushright}
\begin{center}
{\Large\bf Universal Superfield Action for $N=8 \rightarrow N=4$ Partial Breaking
of Global Supersymmetry in $D=1$ }
\end{center}
\vspace{1cm}

\begin{center}
{ S.~Bellucci${}^{a}$, S.~Krivonos${}^{b}$ and A.~Shcherbakov${}^{b}$}
\end{center}

\begin{center}
${}^a$ {\it INFN-Laboratori Nazionali di Frascati, Via E. Fermi 40, 00044 Frascati, Italy}\\
{\tt bellucci@lnf.infn.it} \\\vspace{0.5cm}
${}^b$ {\it Bogoliubov Laboratory of Theoretical Physics, JINR, Dubna, 141980, Russia}\\
{\tt krivonos, shcherb@theor.jinr.ru}
\end{center}
\vspace{1cm}

\begin{abstract}
We explicitly construct $N=4$ worldline supersymmetric minimal off-shell actions for five options of
1/2 partial spontaneous breaking of $N=8, d=1$ Poincar\'e supersymmetry. We demonstrate that the action
for the $N=4$ Goldstone supermultiplet with four fermions and four auxiliary components is a universal one.
The remaining actions for the Goldstone supermultiplets with physical bosons are obtained from the universal one by
off-shell duality transformations.
\end{abstract}

\setcounter{equation}0
\section{Introduction}
Supersymmetric mechanics, being a natural framework for testing the characteristic features it shares with more
complicated higher-dimensional theories, reveals an interesting peculiarity. It turns out that among all
possible $N=4, d=1$ supermultiplets there is a ``root" one \cite{GR}. The action for this ``root"
supermultiplet proved to be a generic one, from which  the actions for the rest of linear and nonlinear $N=4$
supermultiplets can be easily obtained by reduction \cite{root}. This fact seems to be rather important,
and a natural question is whether such universality may show up in other supersymmetric one-dimensional
actions.

Beside the sigma-model type actions there is another very important class of supersymmetric actions -- the one
which describes theories with spontaneous partial breaking of global supersymmetry (PBGS) \cite{hp} - \cite{ik}.
The concept of PBGS provides a manifestly off-shell supersymmetric worldvolume description of various superbranes in terms
of Goldstone superfields \cite{hp}. The physical worldvolume multiplets of the given superbrane are
interpreted as Goldstone superfields realizing the spontaneous breaking of the full brane supersymmetry group down to its unbroken
worldvolume subgroup. The spontaneously broken supersymmetry is realized on the Goldstone superfields by inhomogeneous and nonlinear
transformations.
The choice of the Goldstone supermultiplet is not unique \cite{bg}. Moreover, the proper choice of the Goldstone multiplet
can greatly simplify the construction of the invariant superfield action \cite{{bg},{rt},{gpr}}. In this respect, the preferable
Goldstone supermultiplet should contain the smallest possible number of physical scalars. The reason is rather simple.
The physical scalars in the Goldstone supermultiplets correspond to the central charges in the anticommutators
of manifest and hidden supersymmetry, and these scalars are shifted by constants
under the corresponding transformations. This means that the transformations of the bosonic Goldstone superfields
under hidden supersymmetry contain $\theta$-dependent terms. The presence of such terms makes the construction of the
proper action rather nontrivial.
Moreover, in many cases the invariant Goldstone actions being constructed require
a highly nonlinear redefinition of the Goldstone scalar superfields to bring the action to the standard form (with the Nambu-Goto
action for scalars in the bosonic sector).

In contrast to the higher-dimensional supermultiplets, among the $N=4, d=1$
supermultiplets there is one which does not contain physical bosons at all \cite{{GR},{mult}}. We will use the notation \mult044
to describe this supermultiplet with no physical bosons, four fermions and four auxiliary components.
It is natural to suppose that the
$N=4$ supersymmetric Goldstone superfield action for this supermultiplet with an additional non-linearly realized $N=4$
supersymmetry would give the simplest variant of the system with $N=8\rightarrow N=4$ PBGS.
The aim of this paper is to demonstrate that it is really so. Moreover, we will present the superfield actions
which realize $N=8\rightarrow N=4$ PBGS for all $N=4, d=1$ Goldstone supermultiplets which may be obtained from \mult044
by dualization of the auxiliary components into physical scalars.

\setcounter{equation}0
\section{Universal $N=8\rightarrow N=4$ PBGS action}
Our aim is to construct a $N=4$ superfield action possessing an additional spontaneously broken $N=4$
supersymmetry. It is clear that the $N=4$ superfield formulation is preferable because only $N=4$ supersymmetry
remains unbroken and manifest. We are going to use the \mult044 supermultiplet as a Goldstone one.
So, following \cite{mult}, let us introduce a doublet of fermionic superfields $\Psi_i$ which are subjected to
the constraints
\be\label{sf1}
\bD_i \Psi_j =0,\qquad D^i \bPsi{}^j=0, \qquad D^i \Psi^j + \bD{}^i \bPsi{}^j =0 \;, \qquad i,j=1,2.
\ee
Here, we use the following spinor derivatives:
\begin{equation}
D^{i}=\frac{\partial }{\partial \th_{i}}+ i\bth{}^{i}\partial_{t},\qquad \bD_i=\frac{\partial }{\partial
\bth{}^i}+i\th_{i}\partial _{t},\qquad \left\{ D^i,\bD_j\right\} =2i\delta_j^i\partial_t\;.
\end{equation}
Let us observe that it is immediately follows from \p{sf1} that
\be\label{sf2}
D^2 \Psi^i = -4i \dot{\bPsi}{}^i,\quad \bD{}^2 \bPsi{}^i = 4i \dot\Psi{}^i,
\ee
where
\be
D^2 \equiv D^i D_i , \quad \bD{}^2 \equiv \bD{}^i \bD_i \;.
\ee
In virtue of \p{sf2} the superfield $\Psi_i$ contains among independent components four fermions and four
auxiliary components
\be\label{comp}
\psi_i=\left. \Psi_i\right|_{\theta=0},\; \bpsi_i=\left. \bPsi_i\right|_{\theta=0},\left.
\quad A_{ij}=D_i \Psi_j \right|_{\theta=0}=\left. -\bD_i \bPsi_j \right|_{\theta=0},
\ee
as it should be for an irreducible \mult044 supermultiplet.

As usual, the partial breaking implies the presence of Goldstone fermions among the component fields of the theory.
Assuming that the first components $\psi_i$ \p{comp} of the superfields $\Psi_i$ are just the Goldstone fermions, they should
contain a pure shift in the transformation under spontaneously broken $N=4$ supersymmetry
\be
\delta \Psi^i =\epsilon^i + \ldots ,\quad \delta \bPsi^i =\bar\epsilon^i + \ldots ,
\ee
where $\epsilon^i, \bar\epsilon{}^i$ are the transformation parameters. Clearly, in order to have a linear off-shell realization of
the additional
$N=4$ supersymmetry one has to add one more $N=4$ superfield, but which one? The idea of choosing these additional
superfields is due to J.~Bagger and A.~Galperin \cite{bg} who found that the Lagrangian density for any PBGS action
belongs to an extended supermultiplet. Keeping in mind that any action should start from the free one, which for our
\mult044 supermultiplet reads
\be\label{act1}
S_{free} = \frac{1}{4}\int \! dt d^2 \th\;  \Psi^2 + \frac{1}{4}\int \! dt d^2 \bth \; \bPsi{}^2 \;,
\ee
we will choose this additional $N=4$ superfield $\Phi$ to be a chiral one
\be\label{Phi}
\bD_i \Phi =0 , \quad D^i \bPhi =0 \;.
\ee
So, the proper candidate to be the $N=8 \rightarrow N=4$ PBGS action reads
\be\label{act2}
S = -\int \! dt d^2 \th\;  \Phi + \int \! dt \diff^2 \bth \; \bPhi \;.
\ee
It is easy to find the following transformation laws of $\Psi_i, \bPsi_j$ and $\Phi$ forming the desired $N=4$ supersymmetry algebra:
\bea\label{tr1}
&& \delta \Psi^i = \epsilon^i \left( 1+ \bD{}^2 \bPhi\right) +4i \bar\epsilon{}^i \dot \Phi , \quad
 \delta\bPsi{}^i = \bar\epsilon{}^i \left( 1- D^2 \Phi\right)+4i \epsilon^i \dot \bPhi ,\nn\\
&& \delta\Phi = -\frac{1}{2} \epsilon^i \Psi_i, \quad \delta \bPhi= \frac{1}{2} \bar\epsilon{}^i \bPsi_i,
\eea
with the following Lie bracket:
\begin{equation}\label{closure}
 [\delta_1, \delta_2] = -2i (\eps_2 \bar \eps_1 - \eps_1 \bar \eps_2) {\partial_t} \; .
\end{equation}
It is crucial for us that, in virtue of \p{sf2}, the ``action'' \p{act2} is invariant under \p{tr1}.

Now, in order
to have a meaningful action, one should express the superfields $\Phi$ in terms of our Goldstone
fermionic superfields $\Psi_i$. Following \cite{bg} and motivated by the structure of the free action
\p{act1}, let us start from the following Ansatz:
\be\label{an}
\Phi = \Psi^2 f, \quad \bPhi = - \bPsi{}^2 \bar f ,
\ee
where $f$, $\bar f$ are arbitrary functions depending only on $ \bD^2 \bPhi$ and $D^2 \Phi$, respectively.
Substituting the Ansatz \p{an} in \p{tr1}, one can find that it is consistent, provided
\be\label{sol1}
f = -\frac{1}{4\left( 1+ \bD{}^2 \bPhi\right)},\quad \bar f = - \frac{1}{4\left(1-D^2 \Phi\right)}.
\ee
Thus, with the additional equations
\be\label{eq}
\Phi = -\frac{\Psi^2}{4\left( 1+ \bD{}^2 \bPhi\right)},\quad \bPhi = \frac{\bPsi{}^2}{4\left(1-D^2 \Phi\right)}
\ee
the transformation properties \p{tr1} are satisfied and the action \p{act2} becomes meaningful.

The last step is to solve equations \p{eq}. The procedure simply mimics Bagger-Galperin considerations \cite{bg}
so we omitted the details and will present the answer
\be\label{sol}
\Phi = - \frac{\Psi^2}4 + \bar D^2 \frac{\Psi^2 \bar\Psi^2}{4 \left(1 + \sqrt{1 - 2B} \right)^2}, \quad
 \bar\Phi = \frac{\bar\Psi^2}4 - D^2 \frac{\Psi^2 \bar\Psi^2}{4 \left(1 + \sqrt{1 - 2B} \right)^2},
 \ee
where
\be\label{sol2}
B \equiv D^i\Psi^j  D_i\Psi_j .
\ee
Therefore, with \p{sol} the action \p{act2} acquires the form
\be\label{uniact}
S = \frac{1}{4}\int \! dt d^2 \th\;  \Psi^2 + \frac{1}{4}\int \! dt d^2 \bth \; \bar\Psi^2 -\frac{1}{2}
\int \! dt d^4 \theta \frac{\Psi^2 \bar\Psi^2}{\left(1 + \sqrt{1 - 2D^i\Psi^j  D_i\Psi_j} \right)^2} \;.
\ee
Thus the action \p{uniact} is the desired action which describes a one dimensional system with partially broken
$N=8, d=1$ supersymmetry.

Before going further, let us make some comments. First of all, clearly the supermultiplet $\Psi_i, \Phi$ with the
transformation properties \p{tr1} is a modified version of the $N=8, d=1$ \mult286 multiplet \cite{ABC}.\footnote{
For pedagogical introductions to this subject the reader is invited to consult e.g. \cite{lectures}.}
Secondly, one should stress that the action \p{uniact} describes a very special case of the PBGS theory, because
it does not contain any physical bosonic degrees of freedom. The supermultiplet \mult044 we choose to be
a Goldstone one represents the reduced version of the $N=2,D=3$ double vector supermultiplet \cite{dv} with
all field strengths showing up as auxiliary components. The action \p{uniact} is a one dimensional
version of $N=4 \rightarrow N=2$ PBGS in $D=3$ with the double vector supermultiplet as a Goldstone multiplet.
Finally, one should stress
that the transformation laws \p{tr1} and the action \p{uniact} being so simple in terms of superfields, take a
rather complicated form in terms of components.

\setcounter{equation}0
\section{$N=8\rightarrow N=4$ PBGS superparticle actions}
As in other PBGS theories, the Goldstone fermions can be placed into different multiplets of the unbroken
$N=4, d=1$ supersymmetry. In contrast with the higher-dimensional theories, where each case has to be considered
independently or with using heavily on-/off-shell duality transformations, in one dimension the same
``universal'' action \p{uniact} describes all possible $N=8\rightarrow N=4$ PBGS theories with
different {\it linear} $N=4$ Goldstone supermultiplets. The crucial property for such a universality is the
tight relations between different $N=4$ supermultiplets \cite{{mult},{root}}. In what follows we will present
all possible cases for $N=8\rightarrow N=4$ PBGS.

\subsection{Superparticle in $D=2$}
We start from the simplest situation - the superparticle in $D=2$. Clearly enough, to describe a particle in $D=2$ one
should have one bosonic component in the Goldstone supermultiplet. So, we have to dualize one auxiliary
component in the \mult044 supermultiplet into a physical boson. The resulting $N=4$ Goldstone supermultiplet will be
\mult143, which is defined as \cite{{leva},{mult}}
\begin{equation}\label{c1}
D^i D_i U = \bar D^i \bar D_i U = 0, \quad [D^i,\bar D_i ] U = 0, \qquad \bar U = U.
\end{equation}
If we then identify the spinor derivatives of $U$ and $\Psi$
\begin{equation}\label{c11}
\Psi^i = i\bar D^i U, \quad \bar\Psi^i = i D^i U
\end{equation}
we get exactly the fermionic superfields satisfying (\ref{sf1}). The content of the superfield $U$ includes
one physical bosonic and four fermionic components together with
 a real triplet $A^{ij}=iD^{(i}\bar D^{j)} U|$ of auxiliary fields.

The crucial step is to check whether the transformation law \p{tr1} is compatible with \p{c11}.
It is rather easy to find how
the spontaneously broken $N=4$ symmetry is realized on $U$ and $\Phi$ for this case
\bea\label{1+1}
&& \delta U = -i \epsilon^i (\bar\th_i - 2\bar D_i \bar\Phi) + i \bar\epsilon^i (\th_i - 2 D_i \Phi),\nn \\
&& \delta \Phi = -\frac i2 \epsilon^i \bar D_i U,\quad
\delta \bar\Phi = \frac i2 \bar\epsilon^i  D_i U.
\eea
Upon differentiation, the laws \p{1+1} will reproduce \p{tr1}.
Finally, in order to get the action, one should replace the spinor superfields in the action \p{uniact} by
covariant derivatives of the superfield $U$, according to \p{c11}.

The bosonic part of the action acquire the following form
\begin{equation}
S_{bos} =\int \! dt \left(  1 - \sqrt{1 - 4{\dot u}^2 - 2 A^{ij}A_{ij}}\right),
\end{equation}
which after eliminating the auxiliary fields, turns into
\begin{equation}
S_{bos} = \int \! dt \left(1 - \sqrt{1 - 4{\dot u}^2}\right).
\end{equation}
Thus, the bosonic core is a just Nambu-Goto action in a static gauge for the particle in $D=2$. Therefore the action
\p{uniact}, with the substitution \p{c11}, describes the superparticle with $N=8\rightarrow N=4$ PBGS in $D=2$

\subsection{Superparticle in $D=3$}
In order to get the action of a particle in $D=3$, let us introduce a
twisted chiral multiplet \cite{mult} rather than the standard one
\be\label{twisted}
 D^1 \bLam = D^2 \Lam = 0 , \qquad \bar D_1 \Lam = \bar D_2 \bLam = 0.
 \ee
with \mult242 component structure. The independent components of this supermultiplet may be defined as
\be\label{cont}
\ba{ll}
\mbox{physical bosons:} &  \qquad \lambda, \quad \bar\lambda, \\
\mbox{fermions:} &  \qquad \left. \bar D_1 \bLam\right|, \quad \left. \bar D_2 \Lam\right|, \quad
 \left.   D^1 \Lam\right| , \left. \quad D^2 \bLam\right| , \\
\mbox{auxil. bosons:} &  \qquad \left. A= D^1 \bar D_2 \Lam\right|, \left. {\bar A} =\quad D^2 \bar D_1 \bLam\right| .
\ea
\ee
Now one may immediately check that the spinor superfields $\Psi_i$, $\bPsi^i$ defined as
\be\label{c2}
\Psi_\one = - \bar D_1 \bLam, \quad \Psi_2 = \bar D_2 \Lam, \quad
        \bPsi^\one = D^1 \Lam, \quad \bPsi^2 = - D^2 \bLam
\ee
satisfy the chirality and irreducibility conditions (\ref{sf1}).
The hidden $N=4$ supersymmetry is realized now as follows:
\be\label{1+2}
\ba{l}
\delta \Lam = - \bar\epsilon^1 \th_1 - \epsilon_2 \bar\th^2
    + 2 \, \bar\epsilon^1 D^2 \Phi
    + 2 \, \epsilon_2 \bar D_1 \bar\Phi,\\
\delta \bLam = \epsilon_1 \bar\th^1 + \bar\epsilon^2 \th_2
    + 2 \, \epsilon_1 \bar D_2 \bar\Phi
    + 2 \, \bar\epsilon^2 D^1 \Phi,\\
\delta \Phi = - \frac12 \epsilon_1 \bar D_2 \Lam - \frac12 \epsilon_2 \bar D_1 \bLam, \quad
\delta \bar\Phi = -\frac12 \bar\epsilon^1 D^2 \bLam - \frac12 \bar\epsilon^2 D^1 \Lam.
\ea
\ee

With \p{c2} the bosonic part of the action \p{uniact} acquires the form
\be\label{act3}
S_{bos} = \int dt \left( 1 - \sqrt{1 - 4 \, A \bar A - 16 \dot\lambda \dot{\bar\lambda}} \right).
\ee
The auxiliary fields turn out to vanish on-shell in the bosonic limit, and the static gauge action
of a particle moving in $D=3$ gets the standard form
$$S_{bos} = \int dt \left( 1 - \sqrt{1- 16 \dot\lambda\dot{\bar\lambda} }\, \right). $$
Thus the same superfield action \p{uniact} with the identifications \p{c2} describes the superparticle in $D=3$.

\subsection{Superparticle in $D=4$}
In this case we express the Goldstone superfield $\Psi_i$ through a tensor supermultiplet \cite{341}
\begin{equation}
D^{(i} V^{jk)} = \bar D^{(i} V^{jk)} = 0, \qquad V^{ij} = V^{ji}, \qquad V^{ij}{}^\dagger = V_{ij}
\end{equation}
with \mult341 content. Defining fermionic superfields as
\be\label{c4}
\Psi^i = i\bar D_j V^{ij}, \qquad \bPsi{}^i = i D_j V^{ij}
\ee
one may check that they obey the needed constraints (\ref{sf1}).
The bosonic content of the supermultiplet is a real triplet $v^{ij}=V^{ij}|$ and
a real auxiliary field $M = i  D^i \bar D^j V_{ij}|$.

The spontaneously broken $N=4$ supersymmetry is realized now as
\bea
\delta V^{ij} &=& \frac{2i}3 \left[ \epsilon^{(i} \bar\th^{j)} - \bar\epsilon^{(i}\th^{j)} \right]
    -\frac{4i}3 \left[ \epsilon^{(i} \bar D^{j)}\bar\Phi - \bar\epsilon^{(i} D^{j)} \Phi \right],\\
\delta \Phi &=& \frac{i}{2} \epsilon_i \bar D_j V^{ij}, \qquad \delta \bPhi = - \frac i2\, \bar\epsilon_i D_j V^{ij}.
\eea

With all  these ingredients we may replace the fermionic superfields $\Psi_i$ in the action \p{uniact} by covariant
derivatives of the superfield $V^{ij}$ \p{c4}. After elimination of the auxiliary field the bosonic part of the
action reads
\begin{equation}
S_{bos} = \int dt \left( 1 - \sqrt{\strut 1-18\,{\dot v}^{ij}\,{\dot v}_{ij}}\right),
\end{equation}
and thus it describes the particle in $D=4$.

\subsection{Superparticle in $D=5$}
Finally, we are going to use, as a Goldstone superfield, the $N=4$ ``hypermultiplet''  $Q^{ai}$ defined as
\begin{equation}\label{440}
D^{(i} Q^{j)a} = 0, \qquad \bar D^{(i} Q^{j)a}  = 0, \qquad Q^{ia}{}^\dagger=Q_{ia},\; i,a=1,2.
\end{equation}
This supermultiplet contains four physical bosons and four physical fermions.
Like the previously considered cases, the fermionic superfields constructed as
\be\label{c5}
\Psi_a = i \bar D^j Q_{ja}, \qquad \bPsi_a = i D^j Q_{ja}
\ee
satisfy the relations (\ref{sf1}). The hidden $N=4$ supersymmetry may be realized on the superfields $Q^{ia}, \Phi$ and $\bPhi$ as follows:
\bea\label{delta440}
\delta Q^{ia} &=& \frac i2 (\bar\epsilon^a \th^i - \epsilon^a \bar\th^i )
    + i \epsilon^a \bar D^i \bar \Phi - i \bar \epsilon^a D^i \Phi,\nn\\
\delta\Phi &=& -\frac{i}{2} \epsilon^a  \bar D^i Q_{ia}, \quad \delta \bPhi= \frac{i}{2} \bar\epsilon{}^a  D^i Q_{ia}.
\eea
Once again, the action \p{uniact} with the relations \p{c5} reproduce the action for the superparticle in $D=5$.
Indeed, it is rather easy to find the bosonic core of this action
\begin{equation}
S_{bos} = \int dt \left( 1 - \sqrt{\strut 1 - 32 \, {\dot q}^{ia}{\dot q}_{ia}} \right),
\end{equation}
which describes the particle in $D=5$.

\setcounter{equation}0
\section{Conclusion}
In the present paper we have constructed the universal nonlinear Goldstone superfield action which
is manifestly invariant under $N=4, d=1$ supersymmetry and possesses additional hidden nonlinearly realized
$N=4$ supersymmetry. We have shown that this action, being initially written in terms of the $N=4$ \mult044
supermultiplet, can also serve as the proper action for the rest of the linear $N=4$ supermultiplets. These actions
provide a manifestly world-line supersymmetric description of some superparticles in flat $D=2,...,5$ Minkowski
backgrounds.

We did not plan to give an exhaustive analysis of all possible variants and realizations of partial breaking
of one dimensional $N=8$ supersymmetry. Our goal here was to demonstrate that the universality of some $N=4$
supermultiplets, firstly noted in \cite{{GR},{root}}, may be extended to the case of PBGS actions too.
In contrast with the sigma-model type actions where the ``root'' \mult440 supermultiplet plays the key role, in PBGS
actions the role of ``universal'' supermultiplet is reserved for another supermultiplet -- the \mult044 one. The PBGS
action for this supermultiplet describes some sort of $D$-particle - one dimensional mechanics without
physical bosonic degrees of freedom. Being not too illuminating in itself, this action provides, though, a proper
description of superparticles in various dimensions after dualization of some/or all its auxiliary components into
physical bosons.

In the present paper we limit our consideration by dealing only with linear $N=4$ supermultiplets. But it is
known that there are at least two possible variants of nonlinear $N=4$ supermultiplets \cite{{mult},{HK}}.
The first variant includes two types of $N=4$ nonlinear supermultiplets, which may be obtained from the
nonlinear realization of the $N=4$ superconformal group, in the same manner as the linear ones \cite{mult}.
The second type of nonlinear supermultiplets \cite{HK} is much more complicated and the geometric origin of these
multiplets is still unknown. It is an interesting problem to construct PBGS actions for these types of $N=4$
supermultiplets.

\section*{Acknowledgments}
This research was partially supported by the European Community's Marie Curie Research Training Network under contract
MRTN-CT-2004-005104 Forces Universe, and by the RFBR-06-02-16684 and  DFG 436 Rus 113/669/0-3
grants. S.K. thanks INFN --- Laboratori Nazionali
di Frascati  for the warm hospitality extended to them during the course of this work.

\end{document}